\documentclass[a4paper,10pt]{article}

\usepackage{graphicx,float}

\usepackage[cp1251]{inputenc}

\usepackage{cite,amsmath,amsfonts,amsthm,fullpage}
\usepackage{youngtab}

\usepackage{graphics}
\usepackage{miniplot}
\usepackage{subfigure}

\newcommand{\pb}{\mathbf{p}}

\newcommand{\f}{\textsc{f}}
\newcommand{\e}{\textsc{e}}

\theoremstyle{plain}

\newtheorem{Lemma}{Lemma}
\newtheorem{Proposition}{Proposition}
\newtheorem{Corollary}{Corollary}
\newtheorem{Remark}{Remark}
\newtheorem{Example}{Example}

\theoremstyle{remark}

\def\tr{\mathrm {tr}}
\def\det{\mathrm {det}}

\def\bp{\begin{Proposition}}
\def\ep{\end{Proposition}}
\def\bc{\begin{Corollary}}
\def\ec{\end{Corollary}}
\def\bl{\begin{Lemma}}
\def\el{\end{Lemma}}
\def\be{\begin{equation}}
\def\ee{\end{equation}}
\def\br{\begin{Remark}\rm\small}
\def\er{\end{Remark}}
\def\brs{\begin{remarks}.\\ \rm\
\begin{enumerate}}
\def\ers{\end{enumerate}\end{remarks}}
\def\bea{\begin{eqnarray}}
\def\eea{\end{eqnarray}}

\def\bx{\begin{Example}\rm\small}
\def\ex{\end{Example}}

\def\tr{\mathrm {tr}}
\def\det{\mathrm {det}}

\def\&{&{\hskip -20pt}}

\newcount\YDcount\YDcount=0
\def\YDsize{10pt}

\def\YD#1{%
\ifnum#1=0
 \ifnum\YDcount=0 \ifx\varnothing\undefined\emptyset\else\varnothing\fi
 \else\vskip1.4pt\egroup\YDcount=0\fi
\else
 \ifnum\YDcount=0 \YDcount=1\vcenter\bgroup\vskip1pt
 \else\nointerlineskip\fi
 \vbox{\hrule\hbox{\vrule height\YDsize
 \loop\hskip\YDsize\vrule\ifnum\YDcount<#1\advance\YDcount1\repeat}\hrule
 \kern-0.4pt}\expandafter\YD
\fi}

\usepackage[usenames,dvipsnames]{color}
\usepackage{ulem}

\begin{document}

\author{ Aleksander Yu.
Orlov\thanks{Institute of Oceanology, Nahimovskii Prospekt 36,
Kurchatov Institue,
Moscow, Russia, email: orlovs@ocean.ru }}
\title{Calogero model revisited. Commuting Hamiltonians and Hurwitz numbers}

\date{June 30, 2020}
\maketitle

To Andrey Mironov and Alexei Morozov 

\begin{abstract}
The generalized Mironov-Morozov-Natanson (MMN) equation includes a set of commuting operators, 
which can be considered as Hamiltonians for the quantum Calogero-Sutherland 
problem with a special value of the coupling constant (free fermion point). These Hamiltonians can be 
considered as the center of the enveloping algebra of the group $GL_N(C)$.
Another commuting series of Hamiltonians is presented, parametrized by an arbitrary matrix $A\in GL_N$. 
These Hamiltonians are related to the Hurwitz numbers in the same way as in the case of the MMN equation and 
generate a generalized variant of the Calogero-Surtheland model. 
\end{abstract}

\bigskip

\textbf{Key words:} multivariable commuting differential operators, 
Schur polynomials, Hurwitz numbers, cut-and-join equation, Calogero problem

\textbf{2010 Mathematical Subject Classification:} 05A15, 14N10, 17B80, 35Q51, 35Q53, 35Q55, 37K20, 37K30,

$\,$

\section{Introduction}

The following quantum Hamiltonian
\be\label{cut-and-join-p-alpha}
h(\alpha)= \sum_{n,m>0} p_n p_m p_{-n-m}+p_{n+m}p_{-n}p_{-m}+(\alpha -1)\sum_{l>0} p_ip_{-i}
\ee
independently appears in different areas of mathematical physics \cite{Polychronakos1},\cite{Awata},\cite{OkounkovPand-Calogero},
\cite{Wiegmann},
\cite{SergeevVeselov2009},\cite{SergeevVeselov2013},\cite{Nazarov-Sklyanin2009}. Here $p_n$ is the bosonic 
field which satisfies canonical relation $[p_n,p_m]=\alpha n\delta_{n+m,0}$, $\alpha$ is a parameter. 
The collection  $p_n,\,n>0$ presents creation operators while $p_n,n<0 $ are treated as annihilation ones.
The eigenfunctions of Hamiltonian (\ref{cut-and-join-p-alpha}) turned out to be Jack polynomials 
$J_\lambda^{(\alpha)}(\pb)$ \cite{Awata}. In \cite{Nazarov-Sklyanin2009}  
it was found that (\ref{cut-and-join-p-alpha}) is the Hamiltonian of 
the quantum Benajamin-Ono equation at singular point.

In \cite{GJ} the notion of the cut-and-join equation was introduced, where the differential operator
(\ref{cut-and-join-p-alpha}) with $\alpha=1$ plays an obvious role in splitting $p_{n+m}$ into a product
$p_np_m$ and join it back. This was important because of the correspondence between symmetric functions
and a symmetric group, see \cite{Mac}, where each monomial $\pb_\mu=p_{\mu_1}p_{\mu_2}\cdots$ corresponds to
cycle class $C_\mu$. 
Then Goulden-Jackson cut-and-join equation describes the product $C_{(21^N)}C_\mu$ in 
terms of symmetric functions. 
This equation was used in the problem of enumeration of nonequivalent coverings of the Riemann sphere:
namely it describes the coalence of a branch point with the simple one.
\cite{GJ} pointed out the connection with the KP theory, namely,
experts will explain that $h(1)$ belongs to the symmetries of the KP equation, known as $W_{1+\infty}$ symmetries, 
which were first selected from the famous \cite{DJKM} $\hat{gl}_\infty$ symmetries in \cite{O-1987}.
 In work \cite{Dubr} an interesting observation was made that $h(1)$ can be considered the Hamiltonian of the 
 dispersionless quantum KdV equation. It is not so surprising because the KdV equation is a special limits
 of Benajamin-Ono one. In the future, everywhere $\alpha=1$. This case already has a generalization.

In \cite{MM1},\cite{MMN2},\cite{MM3},\cite{MirMor},\cite{AMMN-2011} the equation which describes the product of
two cycle class was written first in power sum variables and also in the variables $p_n=\tr (X)^n$,
where actual variables are the entries of a matrix $X$, though it can be re-written in terms of eigenvalues
of $X$. I am going to use this point
and consider the generalization of
\be\label{MMN}
W_\Delta(D)\cdot s_\lambda(X)=\tilde{\varphi}_\lambda(\Delta)s_\lambda(X)
\ee
suggested by Mironov, Morozov and Natanzon in \cite{MM1} and called in \cite{MM1} generalized cut-and-join equation,
we will refer it as Mironov, Morozov, Natanzon (MMN) equation.
Here $X$ is a $N\times N$ matrix, $s_\lambda$ is the Schur function (a certain polynomial in the entries of $X$
labeled by a partition (multiindex) $\lambda$. $D$ is the following matrix with entries which
are differential operators
\be
D_{i,j}=\sum_{1\le k \le N} X_{k,j}\frac{\partial}{\partial X_{k,i}}.
\ee
$W_\Delta(D)$ is polynomial in the entries of the matrix $D$ labeled by another partion $\Delta$.
This polynomial is proportional to the {\it normal ordered} power sum polynomial $\pb_\Delta(D)$.
\be
W_\Delta(D)=\frac{1}{\zeta_\Delta}:\pb_\Delta(D):
\ee
where
\be\label{zeta}
\zeta_\Delta=\prod_{i} m_i! i^{m_i}
\ee
where the number $m_k$ indicates how many times 
the part equal to $k$ is included in the partition $\Delta$.

The power sum polynomial  is defined as
\be
\pb_{\Delta}(A)=\tr(A)^{\Delta_1}\cdots\tr(A)^{\Delta_k},\quad \Delta=(\Delta_1,\dots,\Delta_k)
\ee
where $A$ is a matrix. In case we know the eigenvalues of $A$ one can rewrite this polynomial in the entries of $A$ 
as the Newton sum polynomial in the eigenvalues of $A$, we are not going to do it in case $A=D$.
In case $A=D$ is the matrix valued differential operator with the normal ordering

The eigenvalue $\tilde{\varphi}_\Delta(\lambda)$ is a rational number which is proportional to 
the character of the symmetric group in the representation $\lambda$. Detailes will be written down below.

MMN equation describes in a beautiful compact way the result of the coalence of the two branch points with 
arbitrary profiles in the covering problem.
G.Olshanski pointed out that the MMN cut-and-join equation can be treated as the eigenvalue 
problem for a Casimir operator, see \cite{PerelomovPopov1}.

I dedicate this note to Andrei Mironov and Alexei Morozov on the occasion of their 60th birthday and as a token 
of gratitude for their great contribution to the development of mathematical physics in Russia.

\section{Results}

\subsection{Commuting Hamiltonians}

First, instead of $X_{i,j}$ and $\frac{\partial}{\partial X_{i,j}}$ I would rather say that we are dealing with
the algebra of $N^2$ oscillators 
\be\label{pairing}
[Z^\dag_{i,j},Z_{i',j'}]=\delta_{i,i'}\delta_{j,j'},
\quad i,j=1,\dots,N 
\ee
and 
\be
[Z_{i,j},Z_{i',j'}]=0=[Z^\dag_{i,j},Z^\dag_{i',j'}]
\ee

We simply re-denoted matrix $X$ as $Z$ and matrix $D$ as $Z^\dag Z$, where 
$Z^\dag_{i,j}=\frac{\partial}{\partial X_{j,i}}$.
The Fock space $\cal{F}$ is the space of all polynomials in the variables $Z_{i,j},\,i,j=1,\dots,N$.
The left vacuum vector $|0\rangle $ is related to the simplest monomial $1$.
The elements $Z_{i,j}$ and $Z^\dag_{i,j}$ play the role of creation and annihilation operators, respectively.

For a given partition, say, $\mu=(\mu_1,\mu_2,\dots,\mu_k)$ and a matrix $A\in GL_N$ 
we introduce the normal ordered operators (Hamiltonians)
\be\label{Ham}
:\pb_{\mu}(Z^\dag  Z A): = :\prod_{i=1}^{\ell(\mu)} \tr(Z^\dag  Z A)^{\mu_i}  :
\ee
which act on the Fock space $\cal{F}$. 
We use colons ":" as parentheses, indicating that the expression inside them is normally ordered, that
is, all annihilation operators are considered right-shifted relative to the creation operators, for example
$:Z^\dag_{i,j}Z^\dag_{a,b}Z_{j,i}: = Z_{j,i}Z^\dag_{i,j}Z^\dag_{a, b}=:Z^\dag_{a, b}Z_{j,i}Z^\dag_{i,j}:$.
(If we were to use differential operators $\frac{\partial}{\partial X_{j,i}}$ instead of $Z^\dag_{i,j}$, 
we would say that normal ordering means that differential operators do not act on coefficients $
X_{j,i}=Z_{j,i}$ that happen to be inside the normal ordering sign).

\br\label{inside-dots}
Note that under the sign of the normal order, one can permute the traces of operators,
e.g. $:\pb_\mu(Z^\dag ZA)\pb_\nu(Z^\dag Z):\,=\,:\pb_\nu(Z^\dag Z)\pb_\mu(Z ^ \dag ZA):\,$.
We can also, as in the commutative case, rearrange the factors under the trace sign, preserving their cyclic order.
For a given partition, we get $\zeta_\mu$ equivalent ways to write $:\pb_\mu(A):$, where $A$ is an operator-valued
matrix.
\er

Let me remind that the set of commuting operators $\pb_\lambda(Z^\dag Z)$ is actually the set of 
quantum Hamiltonians of the Calogero-Sutherland model at a singular point of the coupling constant 
(free fermionic point) which I call CS $\alpha=1$ model.
and is written in the language of matrix elements $X=Z$, not eigenvalues\footnote{In fact, 
this means that we are dealing not only with the radial part of the Laplace operator $\tr(D)^2$ and of each 
$\pb_\lambda(D)$. It is these Hamiltonians (related to $\alpha=1$ case)
that we will call the Hamiltonians of the Calogero-Sutherland $\alpha=1$ model below.}.

One can prove that for each pair of partitions $\lambda$ and $\mu$ the compatibility condition is satisfied:
\bp
\be
[:\pb_\lambda(Z^\dag Z):,:\pb_\mu(Z^\dag  Z A):]=0
\ee
\ep
Thus, for each individual Young diagram $\mu$, the operator $:\pb_\mu(Z^\dag  Z A):$ is a symmetry of the 
quantum Calogero model.

More interesting is
\bp\label{Prop1} For all partitions $\mu,\nu$ we have
\be\label{commuting}
[:\pb_{\nu}(Z^\dag  Z A):,:\pb_{\mu}(Z^\dag  Z A):]=0,\quad 
\ee
where $Z,A\in GL_N$.
\ep

Proof\footnote{The first version of this text was posted by my mistake, and the proof was not entered there, 
instead of it there was a piece of unfinished text. In this version, only the scheme of the proof will be 
inscribed. The full proof will be posted either separately or in the third version of this text.}.

First of all, note that a product of normal ordered operators can be written as 
a sum of ordered operators, which we write as follows:
\be\label{nu-mu}
:\pb_{\nu}(Z^\dag  Z A): :\pb_{\mu}(Z^\dag   Z A):=\sum_{k=0}^{\min(|\mu|,|\nu|)}:Q_k:
\ee
where the direct analogue of the Leibniz rule is used for the product: $k$ is the number of applied
couplins (\ref{pairing}) of $Z^\dag$ entries of the left factor to the related $Z$ entries  of the right
factor in left hand side of (\ref{nu-mu}) ("related" means
that each $Z^\dag_{ij}$ is coupled to $Z_{ji}$). In particular, we obtain 
$Q_0=\pb_{\mu}(Z^\dag  Z A)\pb_{\nu}(Z^\dag  Z A)$.
In the same way we write
\be\label{mu-nu}
:\pb_{\mu}(Z^\dag  Z A): :\pb_{\nu}(Z^\dag  Z A):=\sum_{k=0}^{\min(|\mu|,|\nu|)}:Q'_k:
\ee 
We see that $Q_0=Q'_0$, see Remark \ref{inside-dots}. Let us show that $Q_k=Q'_k$ for each $k$.

\br
Note that each $\pb_\nu(Z^\dag ZA)$ is a polynomial in the entries of the matrices $Z^\dag, Z, A$ 
and a monomial in these matrices themselves (a monomial in the symbols $Z^\dag$, $Z$, and $C$). 
I note it in order not to create confusion in the use of the terms "monomial" and "polynomial" below: 
"monomial" we are talking about matrices, and "polynomial" is about entries. I also use "monomial in 
entries" as a member of the polynomial.
\er

For clarity, we will place the matrices included in a monomial $\pb_{\lambda_i}$ inside $\pb_\lambda$ on a circle, 
denote it $S^1(\lambda_i)$,
arranging them clockwise in the order in which they are located under the trace sign, if you go from left to right.
So, the get the collection $S^1(\lambda_1),S^1(\lambda_2),\dots$ related to $\pb_\lambda$ and the related
collection of cyclic products - each product is the clockwise product of all matrices around a circle.

Step (i). 

Now we tag a group $k$ of matrices $Z^\dag$ in the monomial $\pb_\nu(Z^\dag Z A)$ (or,
in our vizualized picture - in the collection $S^1(\nu_1),S^1(\nu_n)$), we give these matrices 
numbers and denote them as $Z_{-1},\dots,Z_{-k}$. 
We call these matrices favorite ones.
Each clockwise directed arc which starts at some
favorite $Z_{-i}$ and ends at the nearest clockwise neibroring favorite matrix, say, $Z_{-j}$ we give 
the number $(-i)$ and the related arc product of matrices we denote $C_{-i}$. We suppose that arc 
product does not contain favorite matrices.

We write
$$
:\pb_\nu(Z^\dag ZA): = :\tr (W_1)\cdots \tr(W_n) K_1:
$$
where $n$ ($n\le k$) is the number of circles from the collection above and where each $W_a$
contains at least one favorite matrix. Here $K$ is the product of all factors that do not contain 
favorite matrices. Consider a given $W_a=(Z^\dag_{-i_1}C_{-i_1})
\cdots (Z^\dag_{-i_{f(a)}}C_{-i_{f(a)}})$. Each $C_{-i_p}$ is a product 
$ZA(Z^\dag Z)A)^{n_{i_p}}$ and will be called the arc product, $3n_a+2$ being the length of the arc started at $Z_{-a}$
and ends at the nearest clockwise favorite $Z^\dag$ (note that this nearest favorite $Z^\dag$
can coinside with $Z_{-a}$ itself). 
Note that $Z^\dag$ and $Z$ which enter arc products are not numbered.

Consider a matrix $X\in GL_N$.
we will present each matrix entry $X_{ij}$ as an arrow whose starting point is labeled with $i$ and 
end point is labeled with $j$.  A product of matrices will be drawn as consequent chain of arrows and a 
trace of the product as a closed chain which we will treat as a polygon.

To each $Z_{-a}$ we assign a dotted arrow labeled with $a$ and to each $C_{-a}$ we assign a solid arrow
labeled by $a$. Then we associate each $\tr(W_a)$ with $2f(a)$-gon with alternating dotted and solid edges. 
We call these polygons that emerged from $\pb_\nu(Z^\dag ZA)$ black polygons.

Next we construct "white polygons" from the product $\pb_\mu(Z^\dag ZA)$ and $k$ selected favorite matrices.
We label the group of these $k$-matrices $Z$ in the
monomial $\pb_\mu(Z^\dag ZA)$ (or, equivalently, in the set $S^1(\mu_1),S^ 1(\mu_2),\dots$), where 
we number them somehow and denote them $Z_1,\dots,Z_k$. We consider the set of arcs between clockwise nearest
favorite $Z$. We number an arc by $a$ if it starts at $Z_a$, $a=1,\dots,k$ and directed clockwise. 
We denote related arc products $C_a$.

To each $Z_a$ we associate white dotted arrows, and to each $C_a$ we associate white solid arrows.

 If we pay attention only on favorite matrices we can rewrite $\pb_\mu$ as
 $$
 :\pb_\mu(Z^\dag ZA): = :K_2 \tr(U_1)\cdots \tr(U_m):
 $$
where $K_2$ is the product of all $p_{\mu_i}$ which does not contain any of $Z_a$.
We have $U_a=(Z_{j_1}C_{j_1})
\cdots(Z_{i_h}C_{i_h})$ where each $C_a$ is of form $C_{a}=A(Z^\dag ZA)^{m_{a}})Z^\dag$
with certain $m_a$ and $3m_a+2$ is the length of the arc numbered by $a$.

To each $Z_a$ we associate white dotted arrows, and to each $C_a$ we associate white solid arrows.
Then each is a white polygon with alternating dotted and solid arrows.

For a given selection of favorite matrices we have a set of black polygons emerged from $\pb_nu$
and a set of of white polygons emerged from $\pb_mu$. The coupling of the favorite matrices
$Z_a^\dag$ to $Z_a$ ($a=1,\dots,k$) according to (\ref{pairing}) geometrically means the gluing
of black and white polygons in a way that each black dotted arrow is glued to the oppositely
directed white dotted arrow with the same number.

We get an orientable 2D surface $\Omega$ with the embedded ribbon graph with inflated vertices
Each inflated vertex is a circle (more precisely a negative orientad polygon) obtained as alternating 
black and white solid arrows: the beginning of each solid black arrow is glued to the end of the solid 
white arrow. Let us number the vertices by $i=1,\dots,p$ The cyclic product of matrices around a
vertex $i$ will be denoted $V_i$.
It's Euler characteristic of $\Omega$ is $E=n+m-k+p$.

We have

\bl
By $\langle \cdots \rangle'$ we denote the expectation value only with respect to the matrices
$Z_{\pm a}$, $a=1,\dots,k$ while all other matrices are treated as constant ones which are not coupled.

We get the following relation
\be\label{lemma1}
\frac 1K\langle \pb_\nu(Z^\dag ZA) \pb_\mu(Z^\dag ZA) \rangle'=
\langle \tr(W_1)\cdots \tr(W_n)\tr(U_1)\cdots \tr(U_m) \rangle' =\tr(V_1)\cdots \tr(V_p) 
\ee
where the left hand side is related to glued polygons drawn on a Riemann surface $\Omega$: namely, $ $
where $E=n+m-k+p$ is the Euler characteristic of $\Omega$.The embedded graph drawn on $\Omega$ has $k$
edges, $n+m$ faces (polygons) and $p$ vertices.  Here $V_i$, $i=1,\dots,p$ are cyclic matric products around
the vertex numbered by $i$.
\el

Step (ii).

In the same way we consider the product (\ref{mu-nu}).

Now we  select a set of the favorite matrices $Z^\dag_1,\dots,Z^\dag_k$ in the product $\pb_\mu$ repeating all the steps.
In this case $C_a= ZA(Z^\dag ZA)^{m'_a}$ and $U'_i=(Z^\dag_{i_1}C_{-i_1})\cdots $.
The point that each selected $Z^\dag_{a}$ we find in the white arc numbered by $a$ in step (i).

Next we select a set of the favorite matrices $Z_1,\dots,Z_k$ in the product $\pb_\nu$, where each $Z_a$
is chosen inside
the white arc numbered by $a$. 
In this case $C_a= A(Z^\dag ZA)^{m'_a}Z^\dag$ and $W'_i=(Z_{i_1}C_{i_1})\cdots $.
Then after gluing of obtained polygons we get the same $\Omega$ with the same embedded ribbon graph.
The difference with the case (i) is the different choice of the arc lengths. However the sums of arc lengths
around each inflated vertices of the ribbon graph is the same in both cases.

We get
\bl
\be\label{lemma2}
\frac 1K\langle \pb_\mu(Z^\dag ZA) \pb_\nu(Z^\dag ZA) \rangle''=
\langle \tr(U'_1)\cdots \tr(U'_m)\tr(W'_1)\cdots \tr(W'_n) \rangle'' =\tr(V'_1)\cdots \tr(V'_p) 
\ee
where $\langle \rangle''$ denotes the pairing with respect to the selected matrices and
where $E=n+m-k+p$ is the Euler characteristic of $\Omega$.
\el

Step (iii).

We prove that for each choice of the sets arc lengths $\{n_i,i=1,\dots,n$ and $\{m_i,i=1,\dots,m\}$ 
we find dual sets of $\{n'_i,i=1,\dots,n\}$
and $m'_i,i=1,\dots,m\}$ where right hand sides () and () are equal.

Step (iv).

We prove that $Q_i=Q'_i$ who are obtained after the summation over $k$ and over each selection of the 
favorite matrices.

\paragraph{Symmetric functions}.
Recall that the so-called symmetric functions of variables $x_1,\dots,x_N$ can always be 
expressed in terms of power sum variables $p_n(x_1,\dots,x_N)=\sum_{i=1}^N x^n_i$ (Newtonian sums), see \cite{Mac}.
If $x_1,\dots,x_N$ are eigenvalues of some matrix $X$, then power sums can also be expressed in terms of entries
of this matrix, $p_n=\tr (X)^n$,  and we used just such a representation. What is the analog of  
eigenvalues of a matrix with operator-valued entries I do not know, nevertheless, let's call any 
function in variables $p_i(Z^\dag  Z A),\, i=1,2,\dots$ symmetric function.

We have such an obvious consequence of Proposition \ref{Prop1}

\begin{Corollary} Symmetric functions, understood as functions rewritten in variables of power sums 
$p_n(ZAZ^\dag )$, commute. For example,
 \be
[:\pb_\mu(Z^\dag  Z A):,:s_\lambda(Z^\dag  Z A):]=0,\quad [:s_\lambda(Z^\dag  Z A):,:s_\mu(Z^\dag  Z A) :]=0
 \ee
\end{Corollary}

\subsection{Hurwitz numbers \label{Hurwitz}}

\paragraph{Gluing coverings of the Riemann sphere from polygons.  Hurwitz numbers, geometric approach.}

Let us draw a graph on the Riemann sphere with one vertex ${\cal{O}}$ placed on a 
single edge that divides the surface into two $1$-gones ${\cal{P}}_1$ and ${\cal{P}}_2$.
Let's call this graph $\Gamma$.
We will denote the side of the edge bordering face ${\cal{P}}_1$ with the label $1$, and the other side 
we will denote with the label $-1$.

We can say that the sphere consists of
two $1$-gons ${\cal{P}}_1$ and ${\cal{P}}_2$ whose edges are labeled $1$ and $-1$ respectively
and glued.

Now consider coverings of the sphere of a given degree $D$.

Let's choose points, denote them by 
${\cal{O}}_1$ and ${\cal{O}}_2$ in each $1$-gon and call them the "capitals" of the corresponding $1$-gons
${\cal{P}}_1$ and ${\cal{P}}_2$.
Consider coverings of the sphere with three branch points ${\cal{O}}_1$, ${\cal{O}}_2$, and $\cal{O}$.

Let the point ${\cal{O}}_i$ (where $i=1,2$) has $\ell_i$ pre-images and
let the preimage of 1-gon ${\cal{P}}_i$ with the "capital" ${\cal{O}}_i$ be the set consisting of 
$\mu^{(i)}_1$-, $\mu^{(i)}_2$-,$\dots$ 
$\mu^{(i)}_{\ell_i}$-gons, 
where we agreed that $\mu^{(i)}_1\ge \mu^{(i)}_2 \ge\cdots \ge \mu^{(i)}_{\ell_i} > 0$, i.e., we write 
this set as a partition
$\mu^{(i)}=(\mu^{(i)}_1,\mu^{(i)}_2,\dots,\mu^{(i)}_{\ell_i})$ (or as the Young diagram $\mu^{(i)}$ with line lengths 
$\mu^{(i)}_1,\mu^{(i)}_2,\dots$). 
This partition is called the ramification profile at the point ${\cal{O}}_i$. The set of pre-images of the polygons
${\cal{P}}_i$ we denote 
${\cal{P}}_i(\mu^{i})=\left( {\cal{P}}_i(\mu^{i}_1),\dots, {\cal{P}}_i(\mu^{i}_{\ell_i}) \right)$.

At last let the vertex $\cal{O}$ have $\ell$  pre-images with the valences (i.e. the number 
of outgoing half-edges) $2\nu_1, 2\nu_2,\dots,2\nu_\ell$ which are  on the covering surface, where we the pre-image
of graph $\Gamma$ which we denote $\tilde\Gamma$ is drawn.
Let us assume that $\nu_1\ge \cdots \ge\nu_\ell$. The partition $\nu=(\nu_1,\dots,\nu_l)$ is the ramification profile 
above the point $\cal{O})$. 

Let us note that the weights of partitions $\mu^{(1)}$, $\mu^{(2)}$ and $\nu$ are equal to the degree
of the covering $D$:
\be
|\mu^{(1)}|=|\mu^{(2)}|=|\nu|=D
\ee

Note that the graph $\tilde\Gamma$ is not necessarily simply connected.
Each connected component is a graph drawn on a connected orientable surface. The faces of each component 
consist of pre-images of ${\cal{P}}_1$ and of ${\cal{P}}_2$ and each edge of $\Gamma'$ is a pre-image of the
edge of $\Gamma$,
therefore, it is the boundary between a polygon from the set ${\cal{P}}_1(\mu^{(1)})$ and a polygon from the set 
${\cal{P}}_2(\mu^{(2)})$.
A covering surface can be obtained by gluing  edges of the polygons 
${\cal{P}}_1(\mu^{(1)}_1),{\cal{P}}_1(\mu^{(1)}_2),\dots$) which are pre-images of the side $1$
to edges to the polygons
${\cal{P}}_2(\mu^{(2)}_1),{\cal{P}}_2(\mu^{(2)}_1),\dots$) which are pre=images of the side $-1$.

For given partitions $\mu^{(1)}$ and $\mu^{()2}$ 
there is a finite number of ways of of the  gluings to get a given ramification profile $\nu$.
 Denote this number by ${\cal N}(\mu^{(1)},\mu^{(2)},\nu)$.
We will identify gluings obtained by permuting polygons from the set ${\cal{P}}_i(\mu^{i})$ (here $i=1,2$) 
with the same number of sides or obtained by rotating polygons (cyclic permutation of sides).
If a partition $\mu$ has $m_k$ parts equal to $k$, $k=1,2,\dots$ we introduce the number 
$\zeta_\mu :=\prod_{k=1}^\infty m_k! k^{m_k}$, see (\ref{zeta}), which can be called the order of the automorphisms of 
the Young diagram $\mu$.
The following number 
\be\label{A}
{\cal H}_{\mathbb{S}^2}(\mu^{(1)},\mu^{(2)},\nu)=
\frac{1}{\zeta_{\mu^{(1)}} \zeta_{\mu^{(2)}} } {\cal N}(\mu^{(1)},\mu^{(2)},\nu) 
\ee
is called the Hurwitz number, 
where $\mathbb{S}^2$ point out the base surface. It describes the number 
of non-isomorphic coverings of the sphere with three branch points at which the ramification  profiles 
are given by partitions $\mu^{(1)},\mu^{(2)},\nu$.

It is known that the Hurwitz number does not depend on the order of its arguments. We will reproduce this later.

\paragraph{Wick's rule as gluing rule.} 
If you think a little, then
it is clear that gluing can be represented as a pairing of matrices $Z^\dag$ and $Z$,
\be
\langle (Z^\dag)_{i,j}(Z_{i',j'}\rangle =\delta_{i,i'}\delta_{j,j'},\quad 
i,j=1,\dots,N
\ee
assuming that they are 
assigned to different sides of the edge of $\Gamma$. 
We assign monodromy $W_1=ZC$ to the Northern semisphere and monodromy $W_2=Z^\dag$ to the Southern one.
The monodromy of the vertex of $\Gamma$ is $CF$.
If we drawn a dual graph $\Gamma^*$ 
to graph $\Gamma$, see Figure 1, and assign matrices $Z^\dag$ and $Z$ to half-edges of the dual graph 
and talk about pairing of half-edges according to Leibnitz rule. Constant matrices are assigned to the corners 
of the graph $\Gamma$, that is, to the arcs of the extended vertices, they will also be the arcs of the extended 
vertices of the dual graph $\Gamma^*$ in the Figure 1. For visual illustration: The edge of the dual graph is the 
meridian connecting the North and South Poles (that is, the capitals of the Northern and Southern hemispheres, 
depicted respectively as white and black circles in the Figure 1). 
The collection $(Z^\dag F)^{\mu_1^{(1)}},(Z^\dag F)^{\mu_2^{(1)}},\dots$ describes the monodromy of the 
pre-image of the monodromy $Z^\dag F$ of the southern hemisphere, while the collection
$(ZC)^{\mu_1^{(2)}},(ZC)^{\mu_2^{(2)}},\dots$ describes the monodromy of 
the pre-image of the monodromy $ZC$ of the northern hemisphere.

Wick's rule states that the expectation of the monomial in bosons $Z^\dag_{i,j},Z_{i',j'}$ is calculated 
as follows. We write the monomial as a product of $m$ pairs, this can be done in many ways, namely in $m(2m-1)$ ways, 
if the monomial has $2m$ factors. We consider the product of all pairwise expectations.

Below, we relate polygons ${\cal{P}}_i$ of a graph Gamma to the trace $\tr W_i$ and the components
of the pre-image to $\tr \left(W_i \right)^{\mu^{(i)}_1}$
$$
{\cal{P}}_i \longleftrightarrow \tr W_i,\quad       
 {\cal{P}}_i(\mu^{(i)}_k) \longleftrightarrow   \tr \left(W_i \right)^{\mu^{(i)}_k},
 \quad k=1,\dots, \ell(\mu^{(i)}),\quad i=1,\dots,\textsc{F}
$$
To do it a convenient visualization of this correspondence is as follows. To an entry $C_{i,j}$
(to an entry $F_{i,j}$) where $i,j=1,\dots,N$, we will associate 
a solid white arrow (a solid black arrow), and attribute the beginning of the arrow $i$, 
and the end $j$.  To an entry $Z_{i,j}$ (to an entry $Z^\dag_{i,j}$) 
 we will associate a punctured white arrow (a punctured black arrow), we attribute the beginning of the arrow $i$, 
and the end $j$.

The product of matrices corresponds to a chain of arrows glued to each other, 
and the trace of the product of matrices corresponds to a closed chain: a polygon, which we consider positively 
oriented. The vertices of the polygon contain numbers from $1$ to $N$. The trace of the product is associated 
with the formal sum of such polygons over these numbers.
In case the expressions $\tr W_i$ is described by the formal sum of polygons with, say, $k$ pairs of alternating
solid-white arrors, then $\tr\left( W_i \right)^{\mu_m^{(i)}}$

We get:
\bl\label{L1}
\be
\frac{1}{z_{\mu^{(1)}}z_{\mu^{(2)}}}\langle \pb_{\mu^{(1)}}(Z^\dag F) \pb_{\mu^{(2)}}(Z C)\rangle
=\sum_{\nu} {\cal{H}}_{\mathbb{S}^2}(\mu^{(1)},\mu^{(2)},\nu)\pb_{\nu}(FC),
\ee
where $\quad C,F \in \mathbb{GL}_N(C)$ are independent of $Z$ and $Z^\dag$.
\el
The proof of Lemma \ref{L1} without tudeous details is as follows. As one can see each factor 
in the product $\pb_{\nu_1}(FC)\pb_{\nu_2}(FC)\cdots=\pb_{\nu_1}(FC)$ is related to a 
component of the pre-image of the small face and in this way $\nu$ is the ramification profile of the point 
$\cal{O}$ which is the vertex of the graph $\Gamma_A$. 
While $\pb_{\mu^{(1)}}(Z^\dag F)$ and $ \pb_{\mu^{(2)}}(Z C)$ are related to the pre-images
of polygons ${\cal{P}}_1$ and ${\cal{P}}_1$ respectively and of the capitals ${\cal{O}}_1$ and ${\cal{O}}_2$ 
of these polygons and $\mu^{(1)}$ and $\mu^{(2)}$ are related ramification profiles.

Using (\ref{A}), (\ref{Schur-char-map}) 
and also the orthogonalaty of characters 
\be\label{orth2'}
\sum_\lambda \frac{\dim\lambda}{|\lambda|!} \varphi_\lambda(\mu)\varphi_\lambda(\nu)=\delta_{\mu,\nu}
\ee
one obtains from (\ref{A}) the following:
\bl\label{L-SchurA}
\be\label{SchurA}
\langle s_\lambda(ZC)s_\mu(Z^\dag F) \rangle =\delta_{\lambda,\mu}
\frac{|\lambda|!s_\lambda(C F)}{\dim \lambda},\quad C,F \in \mathbb{GL}_N(C)
\ee
\el

\subsection{Hurwitz numbers and commuting Hamiltonians}

Using Lemma \ref{L1}, the sign of normal ordering and choosing $F=ZA$, we obtain
\bp Let $|\Delta|=|\nu|=n$
\be
:\pb_\Delta (Z^\dag  Z A): |\pb_\nu(ZC)\rangle =
\sum_{\mu\atop |\mu|=n}  |\pb_\mu(ZAC)\rangle \,{\cal{H}}_{S^2}(\Delta,\nu,\mu)
\ee
where ${\cal{H}}_{S^2}(\Delta,\nu,\mu)$ are three-point Hurwitz numbers:
\be
{\cal{H}}_{S^2}(\Delta,\nu,\mu)=
\sum_{\lambda\atop |\lambda|= n} \frac{\dim\lambda}{n!} 
\varphi_\lambda(\Delta) \varphi_\lambda(\nu) \varphi_\lambda(\mu)
\ee
\ep
 
\paragraph{Generalization of MMN cut-and-join equation. } 
 
\begin{Corollary} Suppose $AC=C$ where $A,C\in GL_N$. Then in case $|\Delta|=|\lambda|$ we have
\be\label{genMMN}
:\pb_\Delta (Z^\dag  Z A): |s_\lambda(ZC)\rangle = 
|s_\lambda(ZAC)\rangle \left(\frac{\dim\lambda}{|\lambda|!}\right)^{-1}\chi_\lambda(\mu)
\ee 
\end{Corollary}
 
\paragraph{Related eigenvalue problems}

\begin{Corollary} 
 Suppose $AC=C$ and $|\mu|=|\lambda|$. Then
\be
:\pb_{\mu}(Z^\dag  ZA): s_\lambda(ZC)=
\left(\frac{\dim\lambda}{|\lambda|!}\right)^{-1}\chi_\lambda(\mu) s_\lambda(ZC)
\ee
\end{Corollary}

The important difference with equation (\ref{MMN}) is the restrictive condition $|\lambda|=|\mu|$.
It means that at I present the only set of eigenfunctions as polynomials of the weight $|\mu|$
which is the set $:\pb_\nu(Z^\dag  Z A):|s_\lambda(ZC)\,\rangle $ with all $|\nu|\le \lambda|$.

\section{Another Hamiltonians}

\cite{Regge}, \cite{PolychronakosFeynman}, \cite{PolychronakosReview}

\section{Discussion}

This note is the first part of the work on commuting Hamiltonians and Hutwitz numbers.
It is actually related to the simplest bi-partite embedded graph. 
Fig. 1 corresponds to the simplest bipartite graph with two inflated vertices and disconnected half-edges. 
This graph is dual to the graph considered in Section \ref{Hurwitz}. It clearly shows what the matrices $Z,Z^\dag,F,C$ 
from the Section  correspond to. The black vertex and its half-edge correspond to the Hamiltonian, and the 
white vertex corresponds to the Fock vector.

\begin{figure}[H]
  \centering
  {\includegraphics[width=0.25\textwidth]{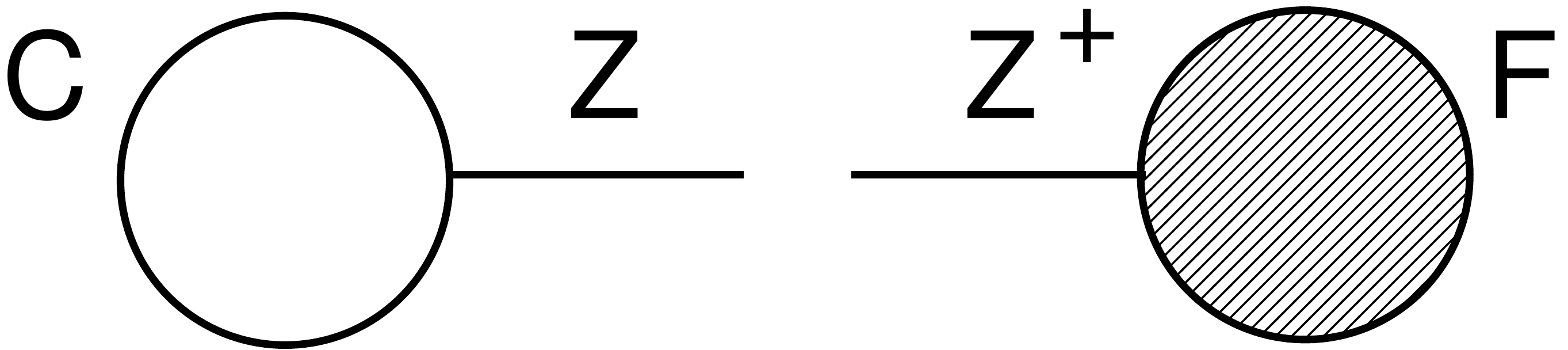}}
  \caption{\label{fig:vertex1}}
\end{figure}

The second part deals with arbitrary embedded bipartite graps.
The simplest example can be presented just now without drawing graphs, it is
\be
:\prod_{i=1}^n \pb_{\mu^i}(Z_iA_i Z^\dag_i): | s_\lambda(Z_1C_1 \cdots Z_nC_n)\,\rangle =
\left(\frac{\dim\lambda}{|\lambda|!}\right) \prod_{i=1}^n \chi_\lambda(\mu^i)
| s_\lambda(Z_1A_1C_1 \cdots Z_nA_nC_n)\,\rangle 
\ee
which is a direct consequence of the above. Actually it is related to the star graph with 
$n$ rays. This is the eigenvalue problem in case $A_iC_i=C_i$.

$\,$

\section*{Acknowledgements}

The author is grateful to  A.D.Mironov, A.Isaev, M. Matushko, 
V.Sokolov and G.Olshanski for stimulating discussions.
He thanks L.Chekhov, N.Slavnov, S.Konstantinous-Rizos, A.E.Mironov for the invitation 
to the workshops in Moscow, Yaroslavl' and Novosibirsk 
where I got useful talks and also Petr and Nadya Grinevich for help.
The work was supported by the Russian Science
Foundation (Grant No.20-12-00195).

\appendix

\section{Partitions and Schur functions \label{Partitions-and-Schur-functions-}}

Let us recall that the characters of the unitary group $\mathbb{U}(N)$ are labeled by partitions
and coincide with the so-called Schur functions \cite{Mac}. 
A partition 
$\lambda=(\lambda_1,\dots,\lambda_n)$ is a set of nonnegative integers $\lambda_i$ which are called
parts of $\lambda$ and which are ordered as $\lambda_i \ge \lambda_{i+1}$. 
The number of non-vanishing parts of $\lambda$ is called the length of the partition $\lambda$, and will be denoted by
 $\ell(\lambda)$. The number $|\lambda|=\sum_i \lambda_i$ is called the weight of $\lambda$. The set of all
 partitions will be denoted by $\mathbb{P}$.

The Schur function labelled by $\lambda$ may be defined as  the following function in variables
$x=(x_1,\dots,x_N)$ :
\be\label{Schur-x}
 s_\lambda(x)=\frac{\det \left[x_j^{\lambda_i-i+N}\right]_{i,j}}{\det \left[x_j^{-i+N}\right]_{i,j}}
 \ee
 in case $\ell(\lambda)\le N$ and vanishes otherwise. One can see that $s_\lambda(x)$ is a symmetric homogeneous 
 polynomial of degree $|\lambda|$ in the variables $x_1,\dots,x_N$, and $\deg x_i=1,\,i=1,\dots,N$.
  
 \br\label{notation} In case the set $x$ is the set of eigenvalues of a matrix $X$, we also write $s_\lambda(X)$ instead
 of $s_\lambda(x)$.
 \er

 There is a different definition of the Schur function as quasi-homogeneous non-symmetric polynomial of degree $|\lambda|$ in 
 other variables, the so-called power sums,
 $\pb =(p_1,p_2,\dots)$, where $\deg p_m = m$.
 
For this purpose let us introduce 
$$
 s_{\{h\}}(\mathbf p)=\det[s_{(h_i+j-N)}(\mathbf p)]_{i,j},
$$
where $\{h\}$ is any set of $N$ integers, and where
the Schur functions $s_{(i)}$ are defined by $e^{\sum_{m>0}\frac 1m p_m z^m}=\sum_{m\ge 0} s_{(i)}(\pb) z^i$.
If we put $h_i=\lambda_i-i+N$, where $N$
is not less than the length of the partition $\lambda$, then
\begin{equation}\label{Schur-t}
 s_\lambda(\mathbf p)= s_{\{h\}}(\mathbf p).
\end{equation}

 The Schur functions defined by (\ref{Schur-x}) and by (\ref{Schur-t}) are equal,  $s_\lambda(\pb)=s_\lambda(x)$, 
 provided the variables $\pb$ and $x$ are related by the power sums relation
  \be
\label{t_m}
  p_m=  \sum_i x_i^m
  \ee
  
  In case the argument of $s_\lambda$ is written as a non-capital fat letter  the definition (\ref{Schur-t}),
  and we imply the definition (\ref{Schur-x}) in case the argument is not fat and non-capital letter, and
  in case the argument is capital letter which denotes a matrix, then it implies the definition (\ref{Schur-x}) with $x=(x_1,\dots,x_N)$ being
  the eigenvalues.

 Relation (\ref{Schur-t}) relates polynomials $s_\lambda$
 and $\pb_\Delta$ of the same degree $d=|\lambda|=|\Delta|$.
 Explicitly one can write
\be\label{powersum-Schur}
\pb_{\Delta}=\sum_{\lambda\in\Upsilon_d} 
\frac{{\rm dim}\lambda}{d!}
\zeta_\Delta\varphi_\lambda(\Delta) s_\lambda(\pb)
\ee
 and
\be\label{Schur-char-map}
s_\lambda(\pb)=\frac{{\rm dim}\lambda}{d!}\sum_{\Delta\in\Upsilon_d } 
\varphi_\lambda(\Delta)\pb_{\Delta}.
\ee
The last relation is called the character map relation.
Here 
\be\label{dim}
 \frac{{\rm dim}\lambda}{d!}: =
\frac{\prod_{i<j\le N}^{}(\lambda_i-\lambda_j-i+j)  }{\prod_{i=1}^{N}(\lambda_i-i+N)!}
\ee
(see example 1 in sect 1 and example 5 in sect 3 of chapt I in \cite{Mac}), where   $N\ge \ell(\lambda)$. As one can check, the right hand side does not depend on $N$.
(We recall that $\lambda_i=0$ in case $i>\ell(\lambda)$. The number
${\rm dim}\lambda$ is an integer.

The factors $\varphi_\lambda(\Delta)$ satisfy the following orthogonality relations
\be\label{orth1}
\zeta_{\Delta}\sum_{\lambda \in\Upsilon_d} 
\left(\frac{{\rm\dim}\lambda}{d!}\right)^2\varphi_\lambda(\mu)\varphi_\lambda(\Delta) =
 \delta_{\Delta,\mu} 
\ee
and
\be\label{orth2}
\left(\frac{{\rm\dim}\lambda}{d!}\right)^2
\sum_{\Delta\in\Upsilon_d} \zeta_\Delta\varphi_\lambda(\Delta)\varphi_\mu(\Delta) =
\delta_{\lambda,\mu}.
\ee
  
\section{Geometrical definition of Hurwitz Numbers \label{Hurwitz-geometric-section}}

In this presentation, we follow article \cite{NO2020}.

The Hurwitz number is a characterisation of the branched covering of a surface with critical values of a 
prescribed topological type.  
Hurwitz numbers of oriented surfaces without boundaries were introduced by Hurwitz
at the end of the 19th century.
Later it turned out that they are closely related to the study of moduli spaces of Riemann
surfaces \cite{ELSV}, to  integrable systems \cite{Okounkov-2000},\cite{Okounkov-Pand-2006},\cite{AMMN-2011} to modern models of mathematical physics
[matrix models], and to closed topological field theories \cite{Dijkgraaf}. In this paper we  consider 
only Hurwitz numbers of compact surfaces without boundary. 

 Consider a branched covering $f:P\rightarrow\Sigma$ of degree $d$ of a compact surface
without boundary. In the neighborhood of each point $z\in P$, the map $f$ is topologically equivalent to the
complex map $u\mapsto u^p$, defined on a neighborhood  $u\sim0$ in $\mathbb{C}$. The number 
$p=p(z)$ is called the degree of the
covering $f$ at the point $z$. The point $z\in P$ is said to be a \textit{branch point} or  
\textit{critical point} if $p(z)\neq 1$. There are only a finite number of critical points. 
The image $f(z)$ of a critical point $z$ is 
called the \textit{critical value} of $f$ at $z$.

Let us associate with a point $s\in\Sigma$ all points  $z_1,\dots,z_\ell\in P$ for which $f(z_i)=s$. Let
$p_1, \dots,p_\ell$ be the degrees of the map $f$ at these points. Their sum $d=p_1 +\dots+p_\ell$ is equal
to the degree $d$ of $f$. Thus, to each point $s\in S $ there corresponds a partition $d=p_1 +\dots+p_\ell$ of
the number $d$. 
Having ordered the degrees $ p_1 \geq \dots \geq p_\ell> 0 $ at each point $ s \in \Sigma $, 
we  introduce the Young diagram $ \Delta^s = [p_1, \dots, p_ \ell ] $ of weight $ d $ with
$ \ell = \ell (\Delta^s) $ rows of length $ p_1 \dots, p_\ell $ :
 $ \Delta^s $ is called the 
\textit{topological type} of the value $ s $, and 
 $ s $ is a critical value of $f$ if and only if at least one of the row-lengths $ p_i $
 is greater than $ 1 $.)

Let us note that the Euler characteristics $\e(P)$ and $\e(\Sigma)$ of the surfaces $P$ and $\Sigma$ are related
via the Riemann-Hurwitz relation:
\[
\e(P)=\e(\Sigma)d +\sum\limits_{z\in P} \left(p(z)-1\right)
\]
or, equivalently,
\be\label{RHur}
\e(P)=\e(\Sigma)d +\sum\limits_{i=1}^\f \left( \ell(\Delta^{s_i})-d\right).
\ee
where $s_1,\dots,s_{\f}$ are critical values.

We say that coverings $ f_1: P_1 \rightarrow \Sigma $ and $ f_2: P_2 \rightarrow \Sigma $ 
are \textit{equivalent}
if there exists a homeomorphism $ F: P_1 \rightarrow P_2 $ such that $ f_1 = f_2F $;
in case $P_1=P_2$ and $f_1 =f_2$ the homeomorphism $F$ is called an
\textit{automorphism of the covering}. 
The set of all automorphisms of a covering $f$ form the group $\texttt{Aut}(f)$
of finite order $|\texttt{Aut}(f)|$. Equivalent coverings have isomorphic automorphism groups.

We present two illustrative examples.

Example 1. Let $ \Sigma = \overline {\mathbb{C}} = \{z \in \mathbb{C} \} \bigcup \infty $,
$ P = P_1 = P_2 = \overline {\mathbb{C}} = \{u \in \mathbb {C} \} \bigcup \infty $  be Riemann spheres.
Consider the branched covering $ z(u) = f (u) = f_1 (u) = f_2 (u) = u^3 $.
This covering $ f: P \to \Sigma $ has 2 critical values 0 and $ \infty $
with Young diagrams from one row of length 3.
Automorphisms of the covering have the form $ F(u) = u^{\sqrt [3] {1}} $.
The group $ \texttt{Aut} (f) $ 
is isomorphic to $ \mathbb{Z} / 3 \mathbb{Z} $.

Example 2. Let $ \Sigma = \overline{\mathbb {C}} = \{z \in \mathbb{C} \} \bigcup \infty $ and
$ P = P_1 = P_2 $ - this is a pair of Riemann spheres; 
that is $ P=P' \bigcup P''$.
where $ P' = \{ u'
\in \mathbb{C} \} \bigcup \infty $ and 
$P'' =\{ u''
 \in \mathbb{C} \} 
\bigcup \infty $.
Consider the branched covering $ z(u') = f (u') = f_1 (u') = f_2 (u') = (u')^3 $,
$ z(u'') = f(u'') = f_1 (u'') = f_2 (u'') = (u'')^3 $.
This covering $ f: P \to \Sigma $ has two critical values 0 and $ \infty $
with Young diagrams of two rows of length 3.
Automorphisms of the covering are generated by the following mappings:

    1. $F(u')=(u')^{\sqrt[3]{1}}$, $F(u'')=u''$.
    
    2. $F(u'')=(u'')^{\sqrt[3]{1}}$, $F(u')=u'$.
    
    3. $F(u')=(u'')$, $F(u'')=u'$.
    
    The group $ \texttt{Aut}(f_i) $ is isomorphic to 
$(\mathbb{Z}/3\mathbb{Z})\bigotimes(\mathbb{Z}/3\mathbb{Z})\bigotimes(\mathbb{Z}/2\mathbb{Z})$.

From now on, unless indicated otherwise, we will assume that the surface $\Sigma$ is connected. Let us choose 
points $ s_1, \dots, s_\f \in \Sigma $ and corresponding Young diagrams $ \Delta^1, \dots, \Delta^\f $ of 
weight $ d $. 
Let $ \Phi $ be the set of  equivalence classes of the coverings for which $ s_1, \dots, s_\f $ is the set 
of all critical values, and $ \Delta^1, \dots, \Delta^\f $ are the topological types of these critical values.
The \textit{Hurwitz number} 
is the number
\be\label{disconH} 
H_{\Sigma}^d(\Delta^1,\dots,\Delta^\f)=\sum_{f\in\Phi} 
\frac {1} {|\texttt{Aut} (f)|}.
\ee
It is easy to prove that the Hurwitz number is independent of the positions of the points $s_1, \dots, s_\f$ 
on $\Sigma$.  One can show that the right hand side of (\ref{disconH})
 depends only on the Young diagrams of $\Delta^1,\dots,\Delta^\f$ and the Euler characteristic $\e=\e(\Sigma)$.
 Because of this sometimes we write $H_{\e(\Sigma)}^d(\Delta^1,\dots,\Delta^\f)$ instead of 
 $H_{\Sigma}^d(\Delta^1,\dots,\Delta^\f)$.

\vspace{1ex}

If $\f=0$ we get an unbranched covering. We denote such Hurwitz number $H_\e\left((1^d)\right)$.

Example 3.
Let $f:\Sigma\rightarrow\mathbb{RP}^2$ be a covering without critical points.
Then, if $\Sigma$ is connected, then $\Sigma=\mathbb{RP}^2$,
$\deg f=1$\quad or $\Sigma=S^2$, $\deg f=2$. Therefore if $d=3$, then
$\Sigma=\mathbb{RP}^2\coprod\mathbb{RP}^2\coprod\mathbb{RP}^2$ or $\Sigma=\mathbb{RP}^2\coprod S^2$.
Thus $H_{1}\left((1^3)\right)=\frac{1}{3!}+\frac{1}{2!}=\frac{2}{3}$.

\section{Combinatorial definition of Hurwitz numbers \label{Hurwitz-combinatorial-section}}

Consider the symmetric group (equivalently, the permutation group) $S_d$ and the equation
\be\label{Hurwitz-combinatorial}
\sigma_1\cdots \sigma_\f \rho_1^2\cdots \rho_\textsc{m}^2\alpha_1\beta_1\alpha_1^{-1}\beta_1^{-1}\cdots
\alpha_{\textsc{h}}\beta_{\textsc{H}}\alpha_{\textsc{h}}^{-1}\beta_{\textsc{h}}^{-1}=1,
\ee
where $\sigma_1,\cdots , \sigma_\f, \rho_1,\cdots, \rho_\textsc{m},\alpha_1,\beta_1,
\dots,\alpha_{\textsc{H}},\beta_{\textsc{h}} \in S_d$, and moreover $\sigma_i \in C_{\Delta^i},\, i=1,\dots,\f$, 
where $C_{\Delta^i}$ is the conjugacy class labeled by a partition $\Delta^i=\left(\Delta^i_1,\Delta^i_2,\dots \right)$.
The Hurwitz number is the number of solutions of equation (\ref{Hurwitz-combinatorial}) divided by $d!$ (by the order
of $S_d$). 

It can be proved that so introduced
the (combinatorial) Hurwitz number coincides with the (geometric) Hurwitz number 
$H_\e(\Delta^1,\dots,\Delta^\f)$ introduced in
Section \ref{Hurwitz-geometric-section}   where  $\e=2-2\textsc{h}-\textsc{m}$.
(One can look at the base surface $\Sigma$ as a result of gluing $\textsc{h}$ handles and $\textsc{m}$ 
M\"obius stripes to a sphere.

Consider the simplest example: $\textsc{h}=0$ and $\textsc{m}=1$; that is $\Sigma=\mathbb{RP}^2$
(real projective plane). Suppose $\f=0$; that is we deal with an unbranched covering. Suppose
$d=3$; that is we consider 3-sheeted covering. Let us solve $\rho^2=1$, where $\rho\in S_3$. One gets
4 solutions: 3 transpositions of the set $1,2,3$ and one identity permutation. There are $3!$
permutations in $S_3$. As a result we get $H_1\left((1^3)\right)=4/3!=2/3$ as we got in the last example
of the previous section.

In the same way one can consider Example 1 of the previous section. In this case $\textsc{h}=\textsc{m}=0$;
that is $\e=2$; one gets the Riemann sphere with two branch points ($\f=2$) and 3-sheeted covering
with profiles 
$\Delta^1=\Delta^2=(3)$. We solve the equation
$\sigma_1\sigma_2=1$, where both $\sigma_{1,2}$ consist of a single cycle of length 3. There are two
solutions $\sigma_1=\sigma_2^{-1}$: one sends $1,2,3$ to $3,1,2$, the other sends $1,2,3$ to $2,3,1$.
We get $H_2\left((3),(3)\right)=2/3!=1/3$.

Example 2 corresponds to $H_2\left((3,3),(3,3) \right)$, $d=6$. One can complete the exercise and get an answer 
$H_2\left((3,3)\right)=1/\zeta_{(3,3)}=1/18$, where $\zeta_\lambda$ is given by (\ref{zeta}). Actually, for any $d$ 
and for any pair of profiles one gets $H_2\left(\Delta^1,\Delta \right)=\delta_{\Delta^1,\Delta}1/\zeta_\Delta$.

In \cite{M2} (and also in \cite{GARETH.A.JONES}) it was found that $H_\e(\Delta^1,\dots,\Delta^\f)$
is given by formula 
\be\label{Mednyh}
H_\e\left(\Delta^1,\dots,\Delta^k \right)=
\sum_{\lambda\in\Upsilon_d}\left(\frac{{\rm dim} \lambda}{d!} \right)^\e
\varphi_\lambda(\Delta^1)\cdots\varphi_\lambda(\Delta^k)
\ee

\section{Differential operators\label{DiffOp} \cite{NO2020F}  }

As G.I.Olshansky pointed out to us, this type of formula appeared in the works of Perelomov and Popov
\cite {PerelomovPopov1}, \cite {PerelomovPopov2}, \cite {PerelomovPopov3}
and describe the actions of the Casimir operators in the representaion $ \lambda $, see
also \cite{Zhelobenko}, Section 9.
 
Here we restrict ourselves only to a reference to important  beautiful works
\cite{Olshanski-19},\cite{Olshanski-199},\cite{Okounkov-199}.

(iv)  Let us notice that if we take a dual graph to the sunflower graph with $n=1$ (dual to one petal $\Gamma$, which is just
a line segment, see fig 1 ), in this case we have one face and two vertices, we get a version of the Capelli-type relation.
Then it is a task to compare explicitly such relations with beautiful results 
 \cite{Okounkov-199},\cite{Okounkov-1996}.

(v)  There are several allusions to the existence of interesting structures related to quantum integrability. 
First, as noted in \cite{NO2020}  by this appearance 2D Yang-Mills theory \cite{Witten}. See also
possible connection to \cite{Gerasimov-Shatashvili}.  Then the appearance of 
the Yangians in works \cite{Olshanski-19},\cite{Olshanski-199} which, we hope, can be related to our subject. 
And finally, the work \cite{Dubr}.

(vi) There is a direct similarity between integrals over complex matrices and integrals over unitary matrices.
However, from our point of view direct anologues of the relations in the present paper are more involved in
the case of unitary matrices. In particular, Hurwitz numbers are replaced by a special combination of these numbers.

\end{document}